\documentclass[twocolumn]{emulateapj}

\shorttitle{Bulge Radial Velocity Assay}
\shortauthors{Howard et al.}

\begin{document}

\title{Kinematics at the Edge of the Galactic Bulge:  Evidence for Cylindrical Rotation}

\author{Christian D. Howard\altaffilmark{1}, R. Michael Rich, Will Clarkson, Ryan Mallery\altaffilmark{1}}
\affil{Department of Physics and Astronomy, UCLA, Los Angeles, CA 90095-1547}
\author{John Kormendy}
\affil{Department of Astronomy, University of Texas at Austin, 1 University Station C1400, Austin, TX 78712-0259, USA}
\author{Roberto De Propris}
\affil{Cerro Tololo Inter-American Observatory, Casilla 603, La Serena, Chile}
\author{Annie C. Robin}
\affil{Observatoire de Besan\c{c}on, Institut Utinam, CNRS-UMR 6213, Universit\'e de Franche-Comt\'e, BP1615, 25010 Besan\c{c}on, France}
\author{Roger Fux}
\affil{Observatoire de Gen\`eve, Universit\'e de Gen\`eve, 51 Ch. des Maillettes, 1290 Sauverny, Switzerland}
\author{David B. Reitzel}
\affil{Griffith Observatory, 2800 East Observatory Road, Los Angeles, CA 90027}
\author{HongSheng Zhao}
\affil{ SUPA, School of Physics and Astronomy, University of St Andrews, KY16 9SS, UK}
\author{Konrad Kuijken}
\affil{Leiden Observatory, Leiden University, PO Box 9513, 2300RA Leiden, The Netherlands}
\author{Andreas Koch}
\affil{Department of Physics and Astronomy, University of Leicester, LE1 7RH, UK}
\altaffiltext{1}{Visiting Astronomers, Cerro Tololo Inter-American Observatory (CTIO).
CTIO is operated by AURA, Inc.\ under contract to the National Science
Foundation.}

\submitted{Received 2009 May 2; accepted 2009 August 6; published 2009 August 24}

\begin{abstract}
We present new results from {\em BRAVA}, a large-scale radial velocity survey of the Galactic bulge, using M giant stars selected from the Two Micron All Sky Survey catalog as targets for the Cerro Tololo Inter-American Observatory 4m Hydra multi-object spectrograph.  The purpose of this survey is to construct a new generation of self-consistent bar models that conform to these observations. We report the dynamics for fields at the edge of the Galactic bulge at latitudes $b=-8^\circ$ and compare to the dynamics at $b=-4^\circ$. We find that the rotation curve {\em V}({\em r}) is the same at $b=-8^\circ$ as at $b=-4^\circ$.  That is, the Galactic boxy bulge rotates cylindrically, as do boxy bulges of other galaxies.  The summed line of sight velocity distribution at  $b=-8^\circ$ is Gaussian, and the binned longitude-velocity plot shows no evidence for either a (disk) population with cold dynamics or for a  (classical bulge) population with hot dynamics.   The observed kinematics are well modeled by an edge-on {\em N}-body bar, in agreement with published structural evidence.  Our kinematic observations indicate that the Galactic bulge is a prototypical product of secular evolution in galaxy disks, in contrast with stellar population results that are most easily understood if major mergers were the dominant formation process.
\end{abstract}


\section{Introduction}

The central bulge of the Milky Way is our nearest example of a spheroidal population, with M31 100 times as distant.  The Galactic bulge stellar population by far is the nearest bulge population that can be studied in star-by-star detail.  Being close enough to permit the study of radial velocity, proper motion, and composition for individual stars, as well as turnoff age for the population,  the bulge/bar population offers us an unprecedented opportunity to test dynamical and formation models for bulge systems.  This is a unique perspective unavailable in the study of extragalactic bulges, which can only be examined from their integrated light, and may change the way we think about the formation of these structures.   The Bulge Radial Velocity Assay ({\em BRAVA})  exploits this unique opportunity with a large scale radial velocity survey of the Galactic bulge/bar population.

The boxy morphology of the central bulge is easily seen in the image of the Galaxy produced by the {\em COBE} satellite (Weiland et al.1994; Dwek et al. 1995; Arendt et al. 1998) and subsequent models have solidified the interpretation that we view an edge-on bar (e.g  Zhao et al. 1996; H{\"a}fner et al. 2000).   Exploiting our proximity,  we know that the bulge/bar is old ($>$10 Gyr) from the modeling of its main-sequence turnoff (Ortolani et al. 1995; Kuijken \& Rich 2002; Zoccali et al. 2003; Picaud \& Robin 2004; Clarkson et al. 2008).  Observations of individual stars at high resolution yields evidence of $\alpha$ enhancement (McWilliam \& Rich 1994; Fulbright, McWilliam, \& Rich 2007; Lecureur et al. 2007) that is modeled to imply a short formation timescale of $<$ 1 Gyr (Ballero et al. 2007).    And while there are hints of an intermediate age population in the bulge at lower Galactic latitudes (van Loon et al. 2003) the bulk of the bulge lacks a convincingly demonstrated intermediate age evolved stellar component (Frogel \& Whitford 1987).  These studies reinforce the widely accepted paradigm that the bulge population was formed both rapidly and early.

Combes \& Sanders (1981) were the first to suggest that galaxy bars heat themselves in the vertical direction and look boxy when seen edge-on.  Additional observations and modeling have confirmed the general picture that boxy and peanut bulges are not spheroidal merger remnants (hereafter "classical bulges") but are in fact edge-on bars (e. g., Combes et al. 1990; Athanassoula \& Misiriotis 2002; Athanassoula 2005; see Kormendy \& Kennicutt 2004 for a review).  However, in our Galaxy, it is difficult to reconcile the early and rapid star formation implied by measurements of bulge star ages and $\alpha$-element overabundances with the protracted star formation that might be expected for a secularly growing bar (Kormendy \& Kennicutt 2004, see Section 8.1).  Freeman (2008) emphasizes that ``in the bar-buckling instability scenario for generating boxy bulges, ... the bulge structure may be younger than its stars, which would originally have been part of the inner disk."  So the old age and rapid star formation history implied by observations of the Galaxy's boxy bulge may not be a problem for our interpretation that this boxy structure was built out of the disk.

Further tests of whether the Galaxy's boxy bulge really is a edge-on bar would therefore be valuable.  Classical bulges are observed to rotate more slowly at increasing height above the disk plane, whereas boxy bulges in other galaxies rotate cylindrically, with $V(r)$ essentially independent of height above the disk plane (e. g., Kormendy \& Illingworth 1982; Jarvis 1990; Shaw 1993; Falc\'on-Barroso et al. 2004; see Kormendy \& Kennicutt 2004 for a review).  Supporting the above interpretation, {\em N}-body models of bars also rotate cylindrically when viewed edge-on (Combes et al. 1990; Fux 1997, 1999; Zhao et al. 1996; Athanassoula \& Misiriotis 2002; Athanassoula 2005). To test our interpretation, we ask: does the boxy bulge of our Galaxy rotate cylindrically or not?

There is also presently some debate as to whether the bar continues to dominate the population at 1 kpc from the bulge.  Zoccali et al. (2008)  suggest that the emergence of a metal poor classical bulge population at higher Galactic latitudes might explain why an abundance gradient is observed in the bulge, as a secularly evolved bar should not have an abundance  gradient.  Further, the bar has been solidly demonstrated at $b=-4^\circ$ (e.g. Stanek et al. 1997) but is not convincingly detected more distant from the plane.  A large sample radial velocity survey at $b=-8^\circ$ offers a sensitive test of whether a classical bulge population is present. 

\begin{figure}[tp]
\centering
\epsscale{1.3}
\plotone{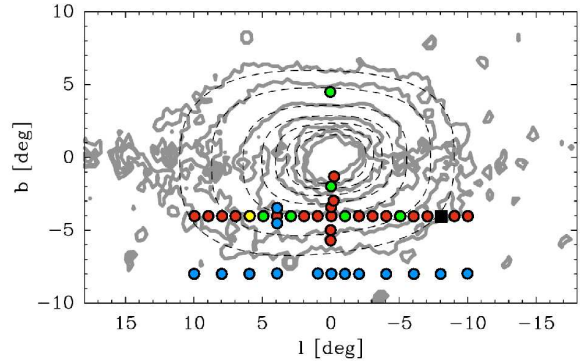} 
\caption{Observed {\em BRAVA} fields up to 2007 August overplotted on the {\em COBE} 2 $\mu$m image (Launhardt et al. 2002).  Green circles represent fields observed in 2005, red circles are fields observed in 2006, and blue circles are fields observed in 2007A ($b=-4^\circ$) and 2007B ($b=-8^\circ$).  The yellow circle is our calibration field, observed with the same fiber configuration every year.  Circle sizes correspond to the 40$^{\prime}$ field of view of the instrument.  The black square is a field observed over multiple years.  For a distance of 8 kpc,  1{$^\circ$} corresponds to 140 pc.
\label{survey}}
\end{figure}

Until recently, we have lacked a kinematic survey of the bulge large enough to seriously test  these bar and bulge dynamical models, and thus influence ideas about bulge evolution.  The kinematic studies of bulge red giant branch (RGB) stars from {\em BRAVA} (Howard et al. 2008) at $b=-4^\circ$ and bulge red-clump giants (Rangwala et al. 2009), show remarkable agreement in the dispersion and rotation curve.    Measurements of the kinematics of the $b=-8^\circ$ strip are especially important, since it gives us an opportunity to investigate whether the bulge rotates cylindrically.   Here we report results from a continuing survey based on red giants, {\em BRAVA}, which comprise the bulk of the $2.4$ $\mu$m light of the bulge.  Early results from this ongoing survey are given in Rich et al. (2007b) and Howard et al. (2008), which report on the 2005-2007A observations.  This Letter reports on the 2007B observations of the $b=-8^\circ$ major-axis strip. 

\section{Observations AND Spectroscopy}
We report on 12 new bulge fields observed in 2007B, sampling the major axis of the bulge in two degree intervals at  $b=-8^\circ$ (Figure \ref{survey}).  We obtained radial velocities for $\sim$100 red giants per field.  We used the Hydra multi-object spectograph at Cerro Tololo Inter-American Observatory (CTIO); the observational setup and target selection were as described in Howard et al. (2008).    Because sources that satisfy our selection criteria are rare at $b=-8^\circ$, we extended our selection limit by $\sim$$0.5$ mag in {\em K} compared to our selection region for the 2005-2007A data (see Figure 9 of Howard et al. 2008).  Although we are observing fainter stars in 2007B than in 2005-2007A, our selection region is still brighter than the ``red-clump" region and samples the same bulge RGB population.  The 2007B radial velocity standards (HD 203638, HD 207076, and HD 218541)  return individual stellar velocities that agree to better than 2 km~s$^{-1}$, on average.  A final velocity was obtained by taking  the error-weighted average of the separate velocity measurements.  One field was observed in all three years (2005, 2006, and 2007B) at $(l,b)=(6^\circ,-4^\circ)$, with the same fiber configuration (i.e., the same stars in that field were observed each year).  Sky conditions during the 2007B night when this field was observed gave less-than-ideal results that are not indicative of the 2007B data in general.  Velocity measurements show a larger rms scatter in velocity differences ($\sim$7 km~s$^{-1}$) as compared to 2005/2006 data, with an average offset of $\sim$2 km~s$^{-1}$ for all stars in that field.  Since the offset is less than the rms scatter, we consider the 2005/2006/2007B data sets to be in good agreement and adopt $\sim$7 km~s$^{-1}$ as our error estimate for individual stellar velocities for all of the data obtained in 2007B.  

\subsection{Color/Magnitude bias}
\begin{figure}[tp]
\centering
\epsscale{1.1}
\plotone{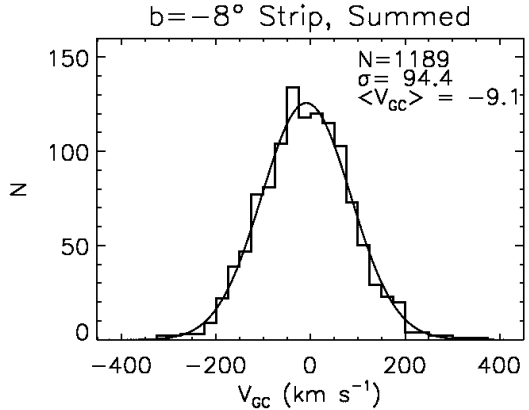}  
\caption{Histogram of all bulge RGB star velocities from the fields at $b=-8^\circ$ ($\sim$1200 stars), in galactocentric velocity ($V_{GC}$).  As with the 2005-2007A data, the co-added sample is consistent with a single kinematic population that is normally distributed and that has negligible skew and kurtosis.  Bin size is 25 km~s$^{-1}$.   
\label{hist}}
\end{figure}

\begin{figure}[tp]
\centering
\epsscale{0.9}
\plotone{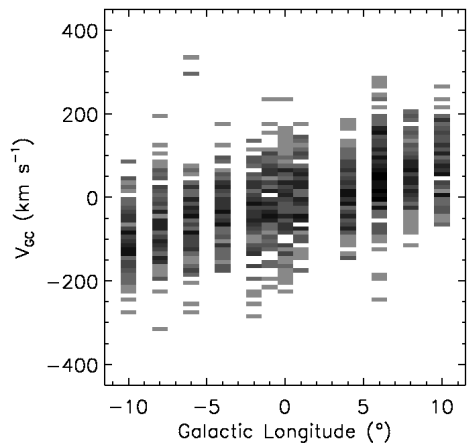} 
\caption{Longitude-velocity ({\em l-v}) plot for the 2007B bulge sample at $b=-8^\circ$; the data are binned to $1^\circ$ in longitude and 10 km~s$^{-1}$ in galactocentric velocity.   There is no evidence for either a cold, rapidly rotating disk, or for a dynamically hot, more slowly rotating (more classical bulge-like) population.   The velocity dispersion drops from that at $-4^\circ$ and is also lower than the halo.    We conclude that the dynamical bar population is predominant at $b=-8^\circ$.
\label{l-v}}
\end{figure}

Observations of the bulge at $b=-8^\circ$ have the potential to probe the bulge/halo boundary.  We examine these $b=-8^\circ$ separately from the higher latitude fields presented in Howard et al. (2008).  To investigate the possibility of color and/or magnitude bias in our sample, the bulge RGB stars from the $b=-8^\circ$ fields are summed and yield an apparent Gaussian distribution, with $<V_{GC}>=-9.1 \pm 2.7 $~km~s$^{-1}$ and  $\sigma=94.4 \pm 1.9$ km~s$^{-1}$ (Figure \ref{hist}).  The longitude-velocity {\em (l-v)} plots for both the $b=-4^\circ$ (see Figure 20 of Howard et al. 2008) and $b=-8^\circ$ (Figure \ref{l-v}) major-axis strips show roughly trapezoidal ``envelopes" that show no hints of contaminating hot (spheroid), or cold (disk), components.  In order to determine if there are color/magnitude biases in our sample, we divide the sample by color and magnitude (see Howard et al 2008 for details) and employ the two-sided Kolmogorov-Smirnov (KS) test.  As with the 2005-2007A data (Howard et al. 2008), we can state that the 2007B color/magnitude-segregated populations are drawn from the same distribution, rejecting the null hypothesis with 96\% confidence.  At $b=-8^\circ$, we are at the point where the bulge density begins to drop sharply; our aim is to search for subpopulations in these regions, and the signature of the inner halo (C. D. Howard et al. 2009, in preparation).  

\section{Evidence of Cylindrical Rotation}
\begin{figure}[tp]
\centering
\plotone{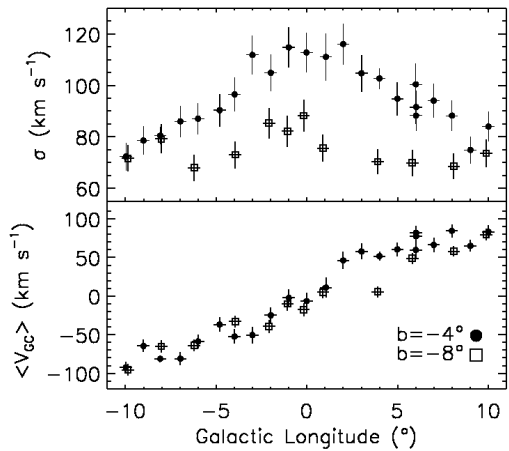}
\caption{Velocity dispersion profile (top) and rotation curve (bottom) for the $b=-4^\circ$ and $-8^\circ$ strips.  The velocity dispersion drops as expected, while the rotation curve is the same for both latitudes, consistent with cylindrical rotation.   The  $b=-8^\circ$ data are also consistent with solid-body rotation, and at first inspection does not appear to show the ÒflatteningÓ observed at $b=-4^\circ$.  More observations are needed to confirm this finding. 
\label{both}}
\end{figure}

We now discuss the Galactic boxy bulge's kinematics at $\sim$560 and $\sim$1120 pc below the disk plane.  Figure \ref{both} shows the dispersion profile and rotation curve for the major axis strips at $b=-4^\circ$ and $-8^\circ$.   Despite the more sparse sampling at $b=-8^\circ$, it is clear that the rotation curve here is indistinguishable from that at $b=-4^\circ$.  We therefore confirm that the Galaxy's bulge rotates cylindrically.  This is further strong evidence that it is an edge-on bar.  It is difficult to conclusively argue whether the rotation curve flattens like that observed at $b=-4^\circ$, however, the inclusion of data obtained at this latitude in 2008 will be able to determine if this is the case (C. D. Howard et al. 2009, in preparation).  

\subsection{Model Comparison}
\begin{figure}[tp]
\centering
\includegraphics[scale=0.3]{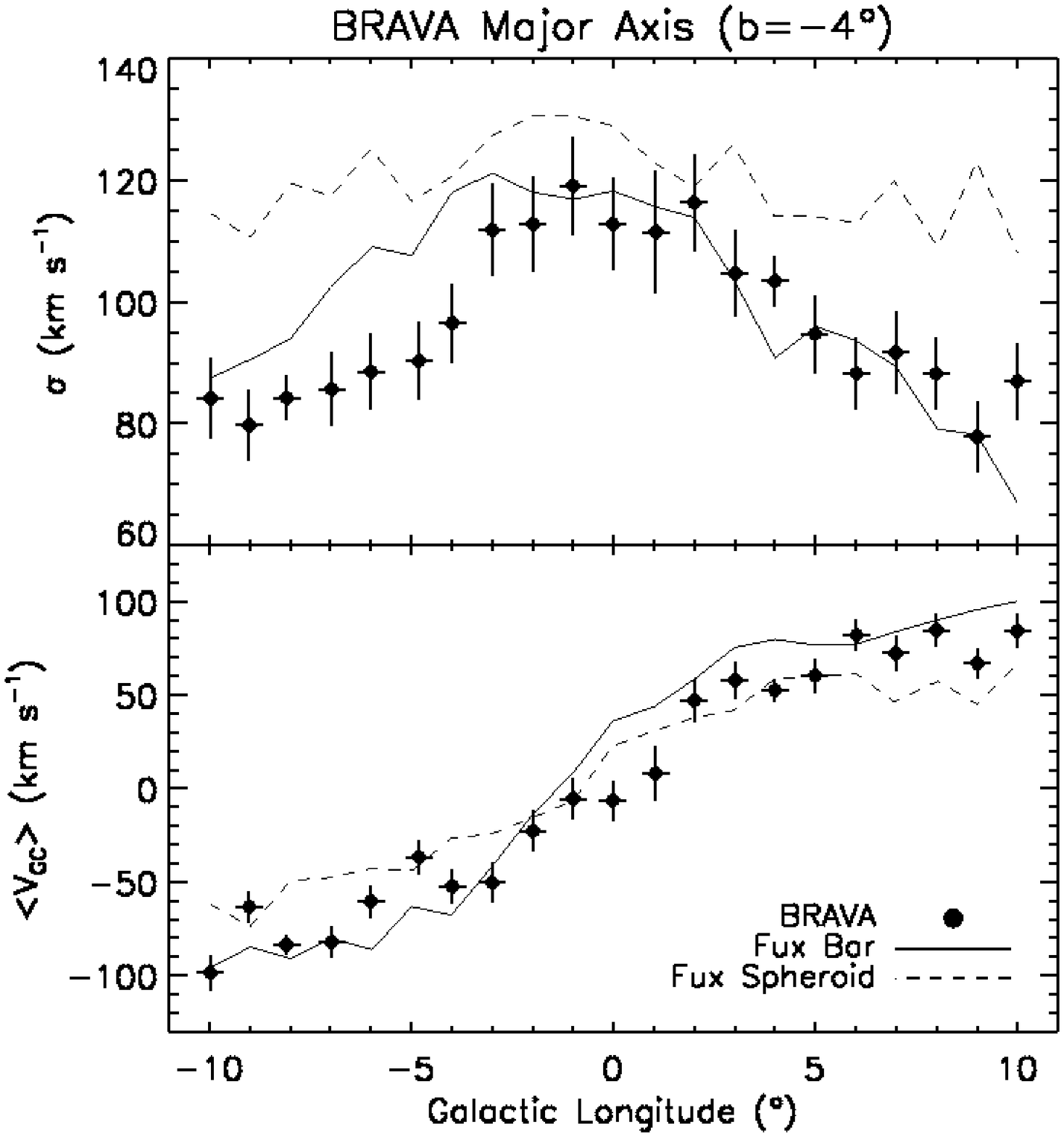}\\
\includegraphics[scale=0.3]{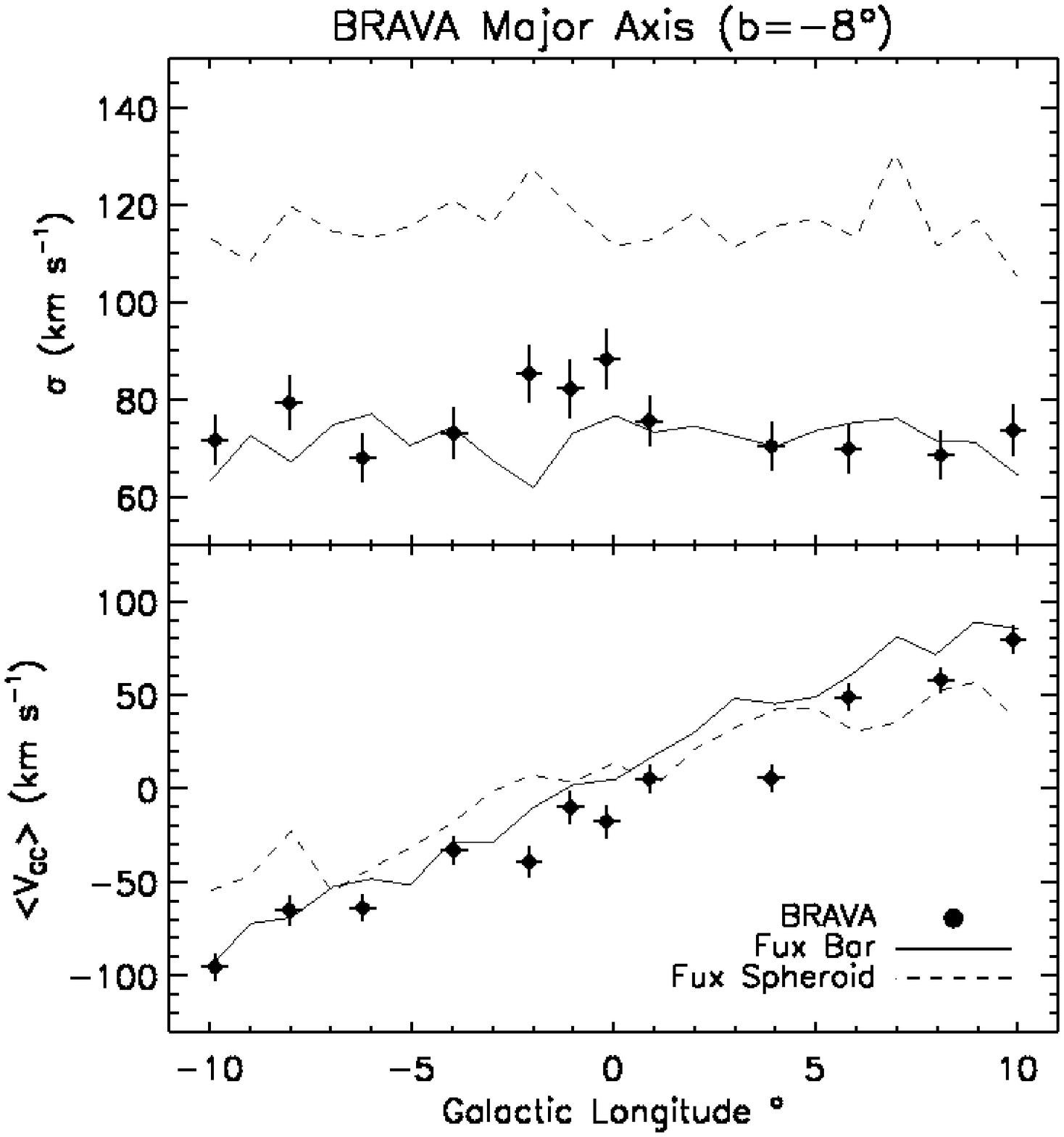} 
\caption{Upper: results for the $b=-4^\circ$ major-axis strip including dispersion (top) and rotation field (bottom), compared with Fux (1999) model components.  Lower: the $b=-8^\circ$  major-axis strip dispersions (top) and mean field velocity (bottom), compared with Fux (1999) model components.   The difference in velocity dispersion between the spheroidal and disk/bar components is striking, and shows that {\em BRAVA} data are well modeled by an {\em N}-body bar.  The rotation speed does not drop with increasing distance from the plane (cylindrical rotation); such cylindrical rotation is a signature of a boxy bulge.
\label{fux}}
\end{figure}

A useful comparison of model and data involves the disentangling of the different stellar components of the model, i.e., spheroid versus disk/bar components, and comparing their kinematics to our observations.  We compare our data to an {\em N}-body bar model (Fux 1999).  The model is constructed from a composite three-dimensional symmetry-free {\em N}-body and hydrodynamics code which follows the constituent particles at a higher resolution than previous models and includes a gas component which reproduces the CO and HI distributions in the ({\em l,b,V}) space (Fux, 1999).  It includes $\sim$$1.5\times10^5$ particles of gas, $\sim$$1.3\times10^6$ particles in the stellar disk/bar, $\sim$$6\times10^5$ particles in a nucleus-spheroid (representing a spheroidal nucleus and a stellar spheroid outside the bar region), and $\sim$$1.7\times10^6$ of dark halo particles.  In Figure \ref{fux}, we compare the {\em BRAVA} data to the disk/bar and nucleus-spheroid components of the model named `c10t2066' described in Fux (1999).  

As one moves further in latitude from the Galactic disk plane, it is evident that the nucleus-spheroid component of the model is hotter and contributes a large velocity dispersion.  Despite this, the {\em BRAVA} data show remarkable agreement with the disk/bar component of the model, with a relatively flat dispersion profile at $\sim$70 km s$^{-1}$ contrasting with the spheroid dispersion of $\sim$120 km s$^{-1}$.  Despite the poorer sampling of fields in 2007B, the data show good agreement with the rotation curve as well, and suggest that the turnover seen in the rotation curve at $b=-4^\circ$ is not evident at $b=-8^\circ$.  Of course, this cannot be confirmed until the rest of the $b=-8^\circ$ major-axis strip data are included. 

\section{Conclusions}

Following the first successful models of the boxy morphology of the $2.4~\mu$m light (Blitz \& Spergel 1991) the case for the deprojected bar has become more convincing (e.g., Bissantz \& Gerhard 2002 models of the {\em COBE} light) and is reinforced by distances of red-clump stars (e.g., Stanek et al. 1997; Babusiaux \& Gilmore 2005).  The boxy outer isophotes and our measurement of cylindrical rotation at $b=-8^\circ$ are also consistent with a bar viewed edge-on; this is one proposed class of  pseudobulge as defined by Kormendy \& Kennicutt (2004).   We emphasize that the rotation curve is strikingly inconsistent with a slowly rotating, dynamically hot, classical bulge population.  The excellent agreement of the disk/bar component of the Fux (1999) {\em N}-body model with the {\em BRAVA} data further supports our interpretation of the Galactic pseudobulge as an edge-on bar.  We assert that the Galactic bulge formation scenario does not adhere to the hierarchical clustering paradigm (which views the bulge as a merger remnant) but rather represents the secular evolution of isolated disks.

A range of {\em N}-body models suggest that the timescale to grow a bar from a massive disk is very short, as few as 2-4 orbital times (Combes et al. 1990; Raha et al. 1991; Sellwood \& Wilkinson 1993, Fux 1997, 1999).   In our Galaxy, this corresponds to $\sim 10^8 $ yr at  1 kpc radius.  Note that other acceleration mechanisms (e.g. vertical resonances; Pfenniger \& Norman 1990) appear to work on longer timescales --- at least $\sim10$ orbital times to achieve vertical thickening.   The observed $\alpha$ enhancements measured for pseudobulge giants (McWilliam \& Rich 1994; Rich \& Origlia 2005, 2007a; Zoccali et al. 2007; Fulbright et al. 2007) would appear to require massive star supernovae, with enrichment proceeding on $\sim10^8$ yr timescales (Ballero et al. 2007).  One may caution that much larger samples of composition measurements, across the entire pseudobulge, are required before the chemical composition becomes a very strong constraint on formation scenarios.  However, present samples of bulge giants are now large enough to permit us to conclude that $\alpha$ enhancement is a characteristic of pseudobulge stars.

\subsection{Discussion}

If the Galactic ``bulge" is a pseudobulge that secularly evolved via dynamical processes, it is difficult to understand how a metallicity gradient, as reported by Zoccali et al. (2008), could arise.  Proctor et al. (2000) find a strong abundance gradient in NGC 4565, which is an edge-on spiral with a boxy bulge.  The Milky Way falls near NGC 4565 on the ({\em V}$_{max}$/$\sigma$) versus $\epsilon$ plot (Howard et al. 2008, Rich et al. 2007b), and the boxy bulge of NGC 4565 also shows cylindrical rotation (Kormendy \& Illingworth 1982).    An abundance gradient in our boxy pseudobulge might arise if there were an old metal-poor spheroid component spatially coincident with the bar.  We argue that Figures 2 and 3 all but rule out a dynamically hot subcomponent that one might assign to such a metal-poor spheroid or classical bulge.   A minor merger might also leave an abundance gradient but would be expected to have left a dynamically cold component behind or possibly large population of evolved intermediate age stars (e.g., carbon stars or luminous M giants as are seen in the LMC bar) if the ingested system had an age range.  No such stellar population is known in the boxy pseudobulge at high Galactic latitudes, and large numbers of carbon stars, for example, would have been easily detected.    An abundance gradient might also arise  if the bar were imprinted rapidly during the violent star-forming phase of the bulge; the relative confinement of metal-rich stars to the plane would argue for some dissipative process accompanying  enrichment.   Soto, Rich, \& Kuijken (2007) find that the bar is most strongly evident in the proper-motion dynamics of the metal-rich population, although the kinematics of stars in our $b=-8^\circ$ fields suggest that virtually all stars in our sample belong to the bar.   

We remain puzzled as to how the boxy pseudobulge could have evolved on a rapid dynamical timescale, yet also have an abundance gradient.  If the stars are formed and are in a massive disk once the buckling process starts, it is difficult to understand how any abundance gradient can be imprinted on the resulting bulge, since the {\em N}-body models accelerate points undistinguished by any physical property.   We believe that our data are now strong enough that one has great difficulty proposing a boxy pseudobulge that transitions into a metal-poor classical bulge at $b\sim-8^\circ$: the required dynamically hot, slowly rotating, population is not observed.  While such a transition would help to explain the gradient, our  {\em BRAVA} selection criteria at $b=-8^\circ$ should include such stars;  the dynamics of the population are remarkably uniform and leave little room for this hypothetical population.  We must conclude, then, that the old population at $b=-8^\circ$ is dominated by the same box/peanut pseudobulge population observed at Galactic latitudes closer to the disk plane.  Even if two types of bulges (both ``classical" and ``box/peanut") coexist in these low latitude fields, it remains problematic as to how there can exist an abundance gradient with such uniform kinematics, but dissipation during the chemical enrichment process offers an interesting route toward a solution.  The {\em BRAVA} survey, however, is finding no indication that a classical bulge population sets in 1 kpc below the plane.

 \acknowledgments  
C.D.H. and R.M.R. acknowledge support from the National Science Foundation, AST-0909479.
J. Kormendy acknowledges support from the National Science Foundation under grant AST-0607490.  C.D.H. acknowledges
support from a CTIO travel grant in support of doctoral thesis observations at CTIO. The authors thank the support staff
at CTIO.  R.M.R. and A.R. acknowledge the hospitality of the Kavli Institute for Theoretical physics at the University of California, Santa Barbara, where some of our discussions took place. This publication makes use of data products from the
Two Micron All Sky Survey, which is a joint project of the University
of Massachusetts and the Infrared Processing and Analysis
Center/California Institute of Technology, funded by the National
Aeronautics and Space Administration and the National Science
Foundation.

{\it Facilities:} \facility{CTIO}


\begin{thebibliography}{}

\bibitem[Arendt et al.(1998)]{1998ApJ...508...74A} Arendt, R.~G., et al.\ 1998, \apj, 508, 74 
\bibitem[Athanassoula \& Misiriotis(2002)]{2002MNRAS.330...35A} Athanassoula, E., \& Misiriotis, A.\ 2002, \mnras, 330, 35
\bibitem[Athanassoula(2005)]{2005MNRAS.358.1477A} Athanassoula, E.\ 2005, \mnras, 358, 1477
\bibitem[Babusiaux \& Gilmore(2005)]{bab05} Babusiaux, C., \& Gilmore, G.\ 2005, \mnras, 358, 1309 
\bibitem[Ballero et al.(2007)]{2007A&A...467..123B} Ballero, S.~K., Matteucci, F., Origlia, L., \& Rich, R.~M.\ 2007, \aap, 467, 123 
\bibitem[Bissantz \& Gerhard(2002)]{2002MNRAS.330..591B} Bissantz, N., \& Gerhard, O.\ 2002, \mnras, 330, 591 
\bibitem[Blitz \& Spergel(1991)]{BS91} Blitz, L., \& Spergel, D.~N.\ 1991, \apj, 379, 631
\bibitem[Clarkson et al.(2008)]{2008ApJ...684.1110C} Clarkson, W., et al.\ 2008, \apj, 684, 1110 
\bibitem[Combes \& Sanders(1981)]{1981A&A....96..164C} Combes, F., \& Sanders, R.~H.\ 1981, \aap, 96, 164
\bibitem[Combes et al.(1990)]{Combes90} Combes, F., Debbasch, F., Friedli, D., \& Pfenniger, D.\ 1990, \aap, 233, 82
\bibitem[Dwek et al.(1995)]{dwek95}Dwek, E. et al. 1995 \apj, 445, 716
\bibitem[Falcon(2004)]{F04} Falc\'on-Barroso, J., et al. 2004, Astron. Nachr., 325, 92
\bibitem[Freeman(2008)]{2008IAUS..245....3F} Freeman, K.~C.\ 2008, IAU Symposium, 245, Formation and Evolution of Galaxy Bulges, ed. Bureau, M., Athanassoula, E., \& Barbuy, B. (Cambridge: Cambridge University Press), 3
\bibitem[Frogel \& Whitford(1987)]{1987ApJ...320..199F} Frogel, J.~A., \& Whitford, A.~E.\ 1987, \apj, 320, 199 
\bibitem[Fulbright et al.(2007)]{Fulbright07} Fulbright, J.~P., McWilliam, A., \& Rich, R.~M.\ 2007, \apj, 661, 1152 
\bibitem[Fux(1997)]{Fux97} Fux, R.\ 1997, \aap, 327, 983 
\bibitem[Fux(1999)]{Fux99} Fux, R.\ 1999, \aap, 345, 787 
\bibitem[H{\"a}fner et al.(2000)]{2000MNRAS.314..433H} H{\"a}fner, R., Evans, N.~W., Dehnen, W., \& Binney, J.\ 2000, \mnras, 314, 433 
\bibitem[Howard et al.(2008)]{2008ApJ...688.1060H} Howard, C.~D., Rich, R.~M., Reitzel, D.~B., Koch, A., De Propris, R., \& Zhao, H.\ 2008, \apj, 688, 1060 
\bibitem[Jarvis(1990)]{J90} Jarvis, B. 1990, in Dynamics and Interactions of Galaxies, ed. R. Wielen (New York: Springer), 416
\bibitem[Kormendy \& Illingworth(1982)]{1982ApJ...256..460K} Kormendy, J., \& Illingworth, G.\ 1982, \apj, 256, 460 
\bibitem[Kormendy \& Kennicutt(2004)]{Kormendy04} Kormendy, J., \& Kennicutt, R.~C., Jr.\ 2004, \araa, 42, 603 
\bibitem[Kuijken \& Rich(2002)]{KR02} Kuijken, K., \& Rich, R.~M.\ 2002, \aj, 124, 2054 
\bibitem[Launhardt et al.(2002)]{Launhardt02} Launhardt, R., Zylka, R., \& Mezger, P.~G.\ 2002, \aap, 384, 112 
\bibitem[Lecureur et al.(2007)]{Lec07} Lecureur, A., Hill, V., Zoccali, M., Barbuy, B., G{\'o}mez, A., Minniti, D., Ortolani, S., \& Renzini, A.\ 2007, \aap, 465, 799
\bibitem[McWilliam \& Rich(1994)]{mr94} McWilliam, A., \& Rich, R.~M.\ 1994, \apjs, 91, 749 
\bibitem[Ortolani et al.(1995)]{Ortolani95} Ortolani, S., Renzini, A., Gilmozzi, R., Marconi, G., Barbuy, B., Bica, E., \& Rich, R.~M.\ 1995, \nat, 377, 701
\bibitem[Pfenniger, D. \& Norman, C.(1990)]{norm90} Pfenniger, D., \& Norman, C. 1990, \apj, 363, 391
\bibitem[Picaud \& Robin(2004)]{2004A&A...428..891P} Picaud, S., \& Robin, A.~C.\ 2004, \aap, 428, 891 
\bibitem[Proctor et al.(2000)]{2000MNRAS.311...37P} Proctor, R.~N., Sansom, A.~E., \& Reid, I.~N.\ 2000, \mnras, 311, 37 
\bibitem[Raha et al.(1991)]{raha91} Raha, N., Sellwood, J.~A., James, R.~A., \& Kahn, F.~D.\ 1991, \nat, 352, 411 
\bibitem[Rangwala et al.(2009)]{2009ApJ...691.1387R} Rangwala, N., Williams, T.~B., \& Stanek, K.~Z.\ 2009, \apj, 691, 1387 
\bibitem[Rich et al.(2007)]{Rich07} Rich, R.~M., Reitzel, D.~B., Howard, C.~D., \& Zhao, H.\ 2007a, \apjl, 658, L29
\bibitem[Rich \& Origlia(2005)]{2005ApJ...634.1293R} Rich, R.~M., \& Origlia, L.\ 2005, \apj, 634, 1293 
\bibitem[Rich et al.(2007)]{ROV07} Rich, R.~M., Origlia, L., \& Valenti, E.\ 2007b, \apjl, 665, L119 
\bibitem[Sellwood \& Wilkinson(1993)]{Sellwood93} Sellwood, J.~A., \& Wilkinson, A.\ 1993, Reports of Progress in Physics, 56, 173 
\bibitem[Shaw(1993)]{Shaw93} Shaw, M. 1993, A\&A, 280, 33
\bibitem[Shaw \& Wilkinson(1993)]{SW93} Shaw, M., Wilkinson, A., \& Carter, D. 1993, A\&A, 268, 511
\bibitem[Stanek et al.(1997)]{stanek97} Stanek, K.~Z., Udalski, A., Szymanski, M., Kaluzny, J., Kubiak, M., Mateo, M., \& Krzeminski, W.\ 1997, \apj, 477, 163 
\bibitem[Soto et al.(2007)]{Soto07} Soto, M., Rich, R.~M., \& Kuijken, K.\ 2007, \apjl, 665, L31
\bibitem[van Loon et al.(2003)]{2003MNRAS.338..857V} van Loon, J.~T., et al.\ 2003, \mnras, 338, 857 
\bibitem[Weiland et al.(1994)]{Weiland94} Weiland, J.~L., et al.\ 1994, \apjl, 425, L81 
\bibitem[Zhao et al.(1996)]{Zhao} Zhao, H., Rich, R.~M., \& Spergel, D.~N.\ 1996, \mnras, 282, 175 
\bibitem[Zoccali et al.(2003)]{zoc03} Zoccali, M., et al.\ 2003, \aap, 399, 931 
\bibitem[Zoccali et al.(2007)]{2007IAUS..241...73Z} Zoccali, M., et al.\ 2007, IAU Symposium, 241, 73 
\bibitem[Zoccali et al.(2008)]{Zocc08} Zoccali, M., Hill, V., Lecureur, A., Barbuy, B., Renzini, A., Minniti, D., G{\'o}mez, A., \& Ortolani, S.\ 2008, \aap, 486, 177 

\end{thebibliography}
\end{document}